\documentclass{svproc}
\usepackage{url}

\usepackage{graphicx}
\usepackage{float}
\usepackage[numbers]{natbib}
\usepackage{amsmath}

\begin{document}
\sloppy
\mainmatter              
\title{Investigating the Fundamental Limit: A Feasibility Study of Hybrid-Neural Archival}
\titlerunning{Investigating the Fundamental Limit}  
%
\author{
    Marcus Armstrong \and
    ZiWei Qiu \and
    Huy Q. Vo \and
    Arjun Mukherjee
}

\institute{
    University of Houston, Houston TX 77004, USA\\
    \email{\{miarmstr, zqiu4, hqvo3\}@cougarnet.uh.edu, arjun@cs.uh.edu}
}

\authorrunning{M. Armstrong et al.}

\maketitle              

\begin{abstract}
Large Language Models (LLMs) possess a theoretical capability to model information 
density far beyond the limits of classical statistical methods (e.g., Lempel-Ziv). 
However, utilizing this capability for lossless compression involves navigating severe 
system constraints, including non-deterministic hardware and prohibitive computational 
costs. In this work, we present an exploratory study into the feasibility of LLM-based 
archival systems. We introduce \textbf{Hybrid-LLM}, a proof-of-concept architecture 
designed to investigate the "entropic capacity" of foundation models in a storage 
context.

\textbf{We identify a critical barrier to deployment:} the "GPU Butterfly Effect," 
where microscopic hardware non-determinism precludes data recovery. We resolve this 
via a novel logit quantization protocol, enabling the rigorous measurement of neural 
compression rates on real-world data. Our experiments reveal a distinct divergence 
between "retrieval-based" density (0.39 BPC on memorized literature) and "predictive" 
density (0.75 BPC on unseen news). While current inference latency ($\approx 2600\times$ 
slower than Zstd) limits immediate deployment to ultra-cold storage, our findings 
demonstrate that LLMs successfully capture semantic redundancy inaccessible to 
classical algorithms, establishing a baseline for future research into semantic file 
systems.

\par\medskip
\noindent\textbf{Keywords:} Neural Compression, Large Language Models, Determinism, Arithmetic Coding, Archival Storage.
\end{abstract}

\section{Introduction}

The rapid expansion of generative AI has created a pressing demand for storage 
density that outpaces the capabilities of traditional entropy coding. Algorithms 
like Deflate and Zstd have asymptotically approached their theoretical limits, 
constrained by their reliance on bounded sliding windows and statistical symbol 
counting \cite{deutsch1996deflate}. Fundamentally, compression is a prediction 
task: the better a model predicts the next token, the fewer bits are required to 
encode it. In this light, Large Language Models (LLMs) offer a paradigm shift: by 
modeling long-range semantic dependencies, they theoretically allow for compression 
rates far exceeding the Shannon limit \cite{6773024} of classical methods. Recent works, 
such as LLMZip \cite{valmeekam2023llmziplosslesstextcompression}, estimate that 
foundational models can achieve compression ratios exceeding 10$\times$ on natural 
language. However, moving from these theoretical bounds to practical 
\textit{Neural Archival} faces two prohibitive system barriers: computational 
latency and hardware non-determinism.

While the density potential is clear, the engineering constraints are severe. 
Standard autoregressive inference scales linearly or quadratically with sequence 
length, making the processing of gigabyte-scale corpora computationally expensive. 
More critically, we identify a stability failure in distributed neural compression 
which we term the \textit{GPU Butterfly Effect}. Due to the non-associativity of 
floating-point arithmetic, parallelized inference (encoding) often yields microscopic 
probability drifts compared to serial inference (decoding). In the context of 
Arithmetic Coding \cite{10.1145/214762.214771,Martin1979RangeEA}, these drifts 
($<10^{-7}$) cause catastrophic decoding divergence, rendering compressed archives 
unrecoverable across heterogeneous hardware.

In this work, we propose \textbf{Hybrid-LLM}, a neural-symbolic architecture 
designed to investigate the feasibility of LLMs as critical storage infrastructure. 
Rather than forcing all data through costly neural pathways, we introduce a 
content-aware \textit{Scout} mechanism that routes low-entropy data to legacy 
compressors, reserving expensive GPU compute strictly for high-density semantic 
compression. To enable rigorous benchmarking, we resolve the determinism crisis via 
a novel logit-quantization protocol, ensuring bit-exact reconstruction regardless of 
the underlying hardware architecture.

This study prioritizes the investigation of "entropic capacity" over immediate 
throughput. We acknowledge that current inference latencies ($\approx 2600\times$ 
slower than Zstd) restrict this approach to ultra-cold archival scenarios. However, 
by establishing a stable, hardware-agnostic testbed, we are able to isolate and 
measure the true compression potential of foundation models.

\textbf{Research Gap \& Organization.} While prior works establish the theoretical 
potential of neural compression, they neglect the systems-level barriers—specifically 
hardware non-determinism—that prevent deployment. The remainder of this paper is 
organized as follows: Section 2 reviews the theoretical foundations of Arithmetic 
Coding. Section 3 details the Hybrid-LLM architecture and the logit-quantization 
protocol. Section 4 presents our empirical evaluation on the Canterbury Corpus and 
the novel December 2025 News dataset. Finally, Section 5 discusses the economic 
implications of the system and concludes with a summary of feasibility.

\textbf{Contributions.} We frame the LLM not merely as a text generator, but as a 
candidate for next-generation entropy coding. Our contributions are:
\begin{itemize}
    \item \textbf{Feasibility Analysis:} We provide a systems-level investigation 
    into the prerequisites for neural archival, identifying hardware determinism and 
    memory bounding as the primary blockers to deployment.
    \item \textbf{Architectural Proof-of-Concept:} We propose a hybrid routing 
    mechanism and a "Grid Snap" logit-quantization protocol that resolve these 
    blockers, enabling the first stable, hardware-agnostic neural compression pipeline.
    \item \textbf{Empirical Limits:} We benchmark the limits of the Llama-3 
    architecture \cite{grattafiori2024llama3herdmodels}, identifying a distinct 
    divergence between "Semantic Deduplication" on memorized text (0.39 BPC) and 
    "Predictive Compression" on novel, unseen data (0.75 BPC).
\end{itemize}

\section{Related Work}

\subsection{Lossless Compression Paradigms}
The field of lossless text compression has historically been dominated by 
dictionary-based and statistical methods. Lempel-Ziv algorithms, such as LZ77 and 
LZ78, form the backbone of widely used standards like GZIP and Zstd 
\cite{deutsch1996deflate,Collet2021Zstandard,duda2014asymmetricnumeralsystemsentropy}, 
effectively exploiting local repetitive structures through sliding windows. While 
these methods are computationally efficient, they are fundamentally limited by their 
inability to capture long-range semantic dependencies, typically plateauing at 
compression ratios around 3-4$\times$ for natural language. More advanced statistical 
approaches, such as the PAQ series (e.g., ZPAQ) \cite{Mahoney2025ZPAQ}, achieve 
significantly higher density by utilizing context mixing. However, these methods 
suffer from extreme computational costs, often requiring hours to process 
gigabyte-scale datasets, which restricts their utility in high-throughput infrastructure.

\subsection{Neural Probability Estimation}
The integration of deep learning with lossless compression shifts the focus from 
symbol counting to \textit{probability estimation}. To understand this paradigm, one 
must view the Large Language Model (LLM) not as a text generator, but as a component 
of Arithmetic Coding (AC).

In AC, the file is represented as a single high-precision floating-point number 
derived from the cumulative probability of its characters. The model's role is to 
narrow the probability interval for the next token based on context. 

\begin{figure}[t]
\centering
\includegraphics[width=\columnwidth]{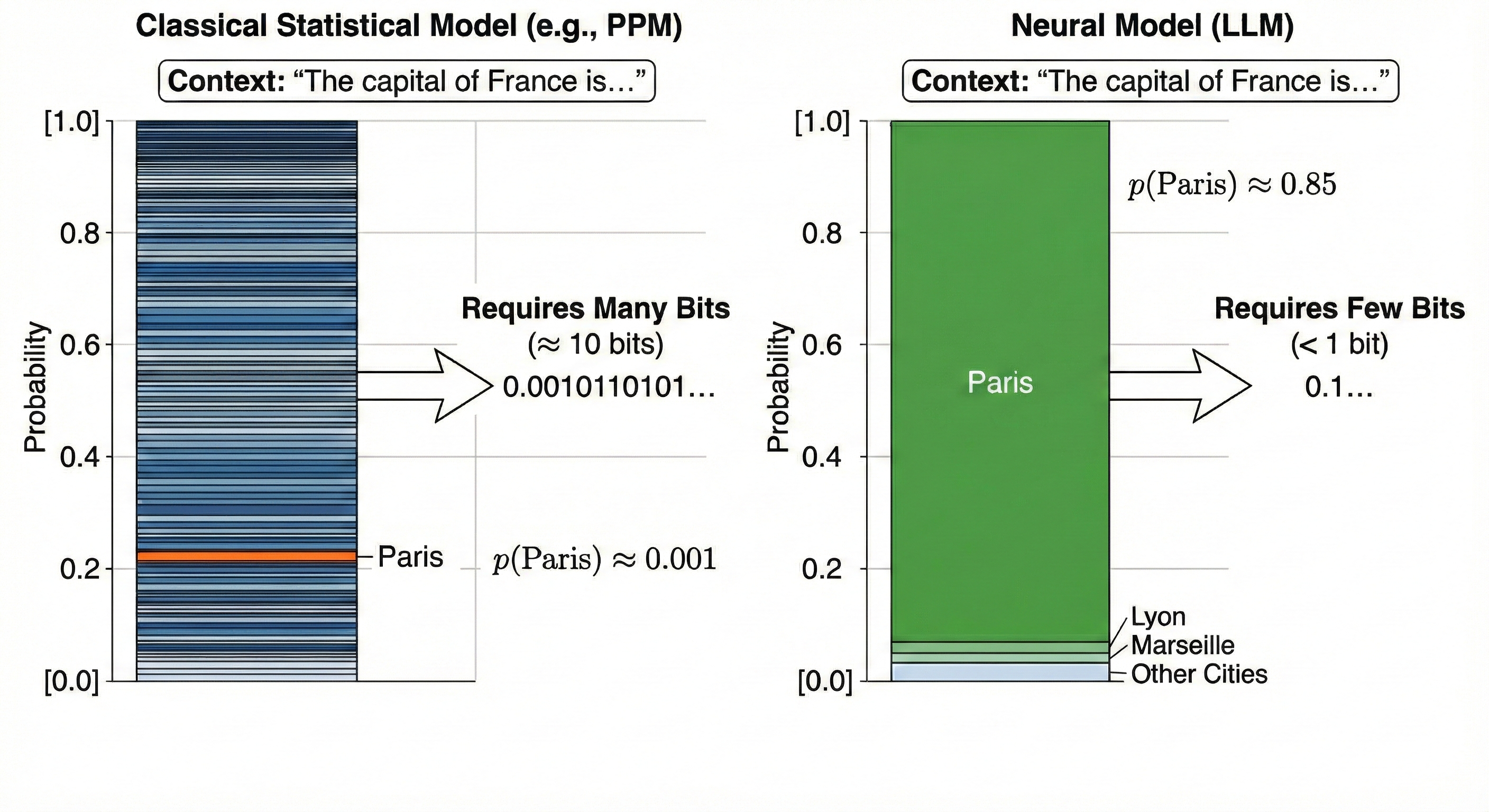}
\caption{Visualizing Neural Arithmetic Coding. Arithmetic coding represents a sequence 
as a precise interval between 0 and 1. (Left) A statistical model, lacking semantic 
understanding, assigns a low probability (narrow interval) to the target word "Paris," 
requiring many bits to define the specific slice. (Right) An LLM leverages context to 
assign a high probability (wide interval) to "Paris." Because wide intervals are easier 
to target numerically, significantly fewer bits are required to encode the data.}
\label{fig:ac_visual}
\end{figure}

As illustrated in \textbf{Figure \ref{fig:ac_visual}}, the efficiency of the compressor 
is directly proportional to the model's confidence. For example, if "Paris" is highly 
likely to follow "The capital of France is...", an LLM assigns it a wide probability 
interval (e.g., $0.0-0.4$). Targeting this wide slice requires fewer bits of precision 
than a statistical model, which might assign it a narrow range ($0.0-0.01$) based on 
simple frequency counts.

Early works, such as DeepZip \cite{goyal2018deepziplosslessdatacompression}, utilized 
Recurrent Neural Networks (RNNs) for this prediction, demonstrating improvements over 
GZIP but struggling with limited context windows. Subsequent approaches like NNCP 
\cite{bellard2019nncp} marked the transition to Transformer-based architectures, 
revolutionizing the domain by enabling highly accurate next-token prediction over 
extended contexts. Modern efforts utilize foundational LLMs 
\cite{touvron2023llamaopenefficientfoundation,touvron2023llama2openfoundation,10.5555/3295222.3295349} 
to maximize this predictive density. The foundational LLMZip study 
\cite{valmeekam2023llmziplosslesstextcompression} formally established that these 
off-the-shelf models could achieve compression ratios exceeding 10$\times$. However, 
LLMZip treated the model as a theoretical "black box," neglecting the systems-level 
necessity of hardware determinism that prevents practical deployment.

\subsection{Determinism and The "Butterfly Effect"}
A critical challenge in neural compression is the non-associativity of floating-point 
arithmetic on GPUs \cite{10.1145/103162.103163}. Arithmetic Coding requires the 
Encoder (Compression) and Decoder (Decompression) to maintain \textit{bit-exact} 
synchronization of probability intervals. If the Encoder (e.g., on an NVIDIA H100) 
calculates a token probability as $0.333333\underline{4}$ and the Decoder 
(on an RTX 4090) calculates $0.333333\underline{3}$, the decoder will select the wrong 
branch of the probability tree.

Recent studies in high-performance computing have confirmed that parallelized 
reduction operations yield non-deterministic results across architectures 
\cite{Nagarajan2018TheIO}. In a compression context, this microscopic drift ($10^{-7}$) 
causes catastrophic failure, rendering the file unrecoverable. Prior works have largely 
evaded this issue by running experiments in strictly controlled, single-hardware 
environments. In contrast, our work directly addresses this "Butterfly Effect" through 
explicit logit quantization, enabling the first robust, distributed neural compression 
system that guarantees decoding correctness across heterogeneous compute environments.

\section{Methodology}

\begin{figure}[t]
\centering
\includegraphics[width=\columnwidth]{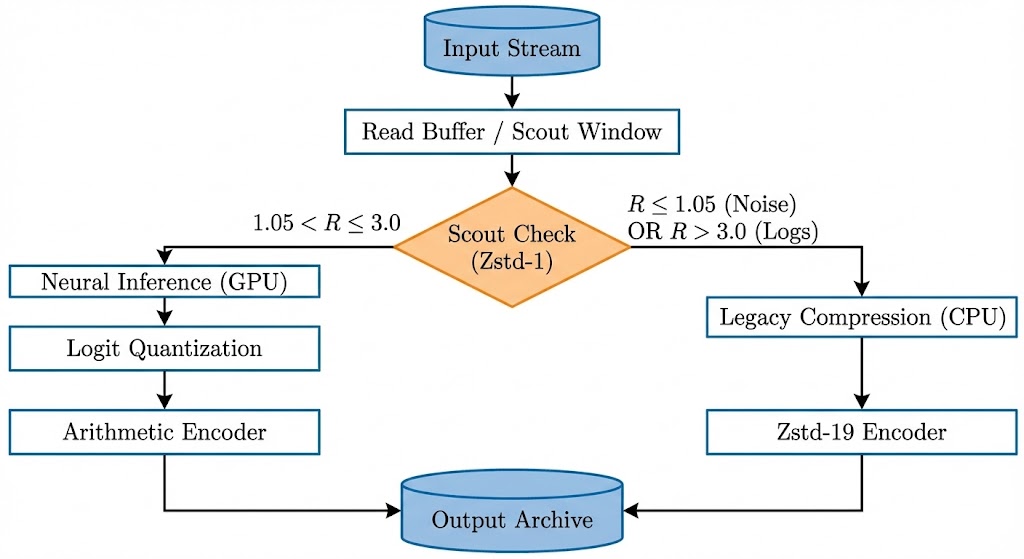}
\caption{Content-Aware Hybrid Routing Logic. The system processes the input stream 
in segments. A lightweight 'Scout' (Zstd-1) filters out incompressible noise ($R \le 1.05$) and 
highly redundant logs ($R > 3.0$) to the CPU. Only data in the 'Semantic Zone' is routed to the 
GPU, ensuring that expensive neural inference is reserved for data where it yields information gain.}
\label{fig:routing_logic}
\end{figure}

We propose \textbf{Hybrid-LLM}, a comprehensive architecture designed to transform Large 
Language Models from theoretical predictors into practical, critical infrastructure for data 
storage. Our methodology addresses the three fundamental barriers identified in our feasibility analysis: 
computational inefficiency on low-entropy data, hardware non-determinism in distributed 
environments, and the latency of autoregressive inference. By synergizing classical 
compression techniques with advanced neural mechanisms, we establish a robust pipeline that 
balances density, speed, and correctness.

\subsection{Mathematical Formulation}
Let $S = \{x_1, x_2, \dots, x_N\}$ be a sequence of tokens. The theoretical description 
length $L(S)$ under an optimal code is given by the negative log-likelihood of the sequence 
under the model $P_\theta$:
\begin{equation}
    L(S) = -\sum_{t=1}^{N} \log_2 P_\theta(x_t | x_{<t})
\end{equation}
In standard implementations, $P_\theta$ represents the raw Softmax output of the LLM. However, 
we identify that $P_\theta$ is hardware-dependent due to non-associative floating-point 
summation. To address this, we introduce a quantized probability distribution $Q_\phi$, 
defined below.

\subsection{Content-Aware Hybrid Routing}
The primary inefficiency of prior neural compression attempts lies in 
their monolithic treatment of data; they apply computationally expensive 
inference to every byte, regardless of its semantic complexity. We posit 
that neural intelligence is a scarce resource that should be allocated 
only where it yields significant information gain. To realize this, we 
introduce a lightweight "Scout" mechanism (see \textbf{Figure \ref{fig:routing_logic}}) utilizing Zstandard (Zstd) at 
level 1 \cite{Collet2021Zstandard} to pre-scan incoming data blocks. This scout acts as a low-latency 
estimator of entropic complexity before the data reaches the neural engine.

Let $R_{zstd}(b)$ be the compression ratio of block $b$ using the Scout. 
We implement a "Two-Pass, Band-Pass" routing strategy. If a block is 
routed to the CPU, it is re-compressed using Zstd at level 19 to maximize 
density. While this implies a second pass over the data, the computational 
cost of classical compression (measured in microseconds) is negligible 
compared to the latency of neural inference (measured in milliseconds), 
rendering the overhead statistically insignificant relative to the 
system's total runtime.

To prevent GPU wastage, we define an ideal zone for neural 
processing. We recognize that while LLMs excel at compressing semantic 
structure, they perform poorly on high-entropy randomness 
(such as encrypted archives) where the compression ratio approaches unity. 
Routing such noise to the GPU would incur the maximum compute cost for 
near-zero gain. Therefore, our routing function $f(b)$ incorporates both a 
lower and upper bound:
\begin{equation}
    f(b) = 
    \begin{cases} 
      \text{GPU\_Encode}(b) & \text{if } \tau_{min} < R_{zstd}(b) \leq \tau_{max} \\
      \text{CPU\_Encode}(b) & \text{otherwise}
    \end{cases}
\end{equation}
We empirically set the thresholds at $\tau_{min} = 1.05$ and 
$\tau_{max} = 3.0$ to segregate data into three distinct entropic 
regimes. Data falling below the lower threshold ($R \leq 1.05$) represents 
incompressible noise or encryption, which is routed to the CPU to avoid 
wasting GPU cycles. Conversely, data exceeding the upper threshold 
($R > 3.0$) typically indicates high syntactic redundancy, such as 
structured logs or zero-padding, which Zstd captures efficiently without 
the need for neural prediction. The Neural Engine is therefore reserved 
exclusively for the intermediate regime ($1.05 < R \leq 3.0$), where the 
data exhibits complex semantic dependencies that classical dictionaries 
fail to model effectively. This content-aware routing creates a symbiotic 
relationship between AI and classical algorithms, ensuring that the heavy 
energy cost of the LLM is incurred only when it provides a distinct 
advantage over standard tools.

\subsection{Deterministic Neural Coding via Logit Quantization}
\begin{figure}[t]
\centering
\includegraphics[width=\columnwidth]{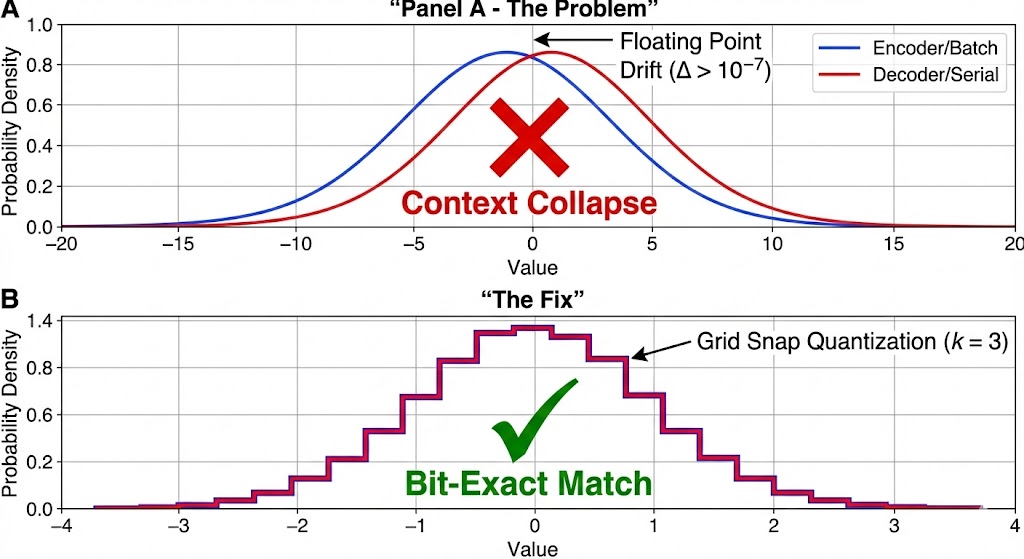}
\caption{Visualizing the GPU Butterfly Effect. (Top) Without quantization, 
microscopic floating-point drifts between parallel encoding and serial decoding accumulate, 
causing the arithmetic coder to diverge. (Bottom) Our protocol quantizes logits 
into a discrete probability space, enforcing bit-exact reproducibility across heterogeneous 
hardware.}
\label{fig:determinism}
\end{figure}

For neural compression to serve as critical infrastructure, the decoding 
process must be bit-exact reproducible. However, we identify two distinct 
failure modes in distributed environments, visualized in \textbf{Figure \ref{fig:determinism}}: the non-associativity of 
floating-point accumulation (the "Butterfly Effect") and the 
hardware-dependence of transcendental functions. Parallelized matrix 
multiplication kernels on GPUs often produce output logits that drift by 
magnitudes of $10^{-7}$ \cite{micikevicius2018mixedprecisiontraining} compared to serial kernels. Furthermore, standard 
hardware implementations of the exponential function ($\exp$) are not 
guaranteed to be bit-exact across different architectures 
(e.g., NVIDIA vs. Intel), as they are not standardized by IEEE 754 to the 
same degree as basic arithmetic. In arithmetic coding, where the entire 
message is encoded into a single interval \cite{5391119}, these microscopic drifts 
desynchronize the sender and receiver, leading to catastrophic context 
collapse.

To enforce strict determinism, we introduce a 
\textit{Logit Quantization and Host-Offload Protocol}. We strictly isolate 
the non-deterministic GPU operations from the sensitive probability 
construction. First, raw logits $z_t$ are quantized in-place on the GPU to 
a coarse precision $k=3$ decimal places:
\begin{equation}
    \hat{z}_t = \frac{\lfloor z_t \cdot 10^k \rceil}{10^k}
\end{equation}
This method effectively filters out the accumulation noise inherent 
to deep transformer layers. Second, to mitigate transcendental drift, 
these quantized logits are transferred to the host CPU and cast to 
IEEE 754 double-precision (\texttt{float64}). The final coding 
distribution $Q_\phi$ and the resulting Cumulative Distribution 
Function (CDF) are computed strictly on the host using standard libraries:
\begin{equation}
    Q_\phi(x_t | x_{<t}) = \frac{\exp(\hat{z}_t)}{\sum_{j=1}^{|V|} \exp(\hat{z}_j)}
\end{equation}
By offloading the Softmax operation, we ensure that the transcendental 
approximations are consistent regardless of the accelerator used for 
inference. We prefer quantizing in logit-space rather than 
probability-space because the output distribution of an LLM is typically 
long-tailed; direct quantization of probabilities risks zeroing out the 
mass of rare tokens, which would break the arithmetic coder. Quantizing 
logits preserves the relative ordering of the tail while stabilizing the
numerical substrate. We empirically confirmed that the entropy penalty of 
this protocol is negligible ($< 0.002$ BPC) on the \textit{Alice} corpus.

\textbf{Quantization Penalty Analysis.} A concern with coarse quantization 
($k=3$ decimal places) is the potential loss of compression density due to increased 
Kullback-Leibler (KL) divergence \cite{cover2006elements} from the true distribution. 
However, empirically, we observe a bit-rate penalty of less than $0.002$ BPC. This 
stability arises because arithmetic coding is most sensitive to the probabilities of 
high-frequency tokens. In a standard LLM vocabulary (approx. 128k tokens), the 
probability mass is concentrated in the top-$k$ logits. Our protocol preserves the 
relative ranking and magnitude of these dominant tokens, while the rounding noise 
primarily affects the low-probability tail, which contributes negligibly to the 
overall code length.

\subsection{Bounded-Memory Block Processing}
\begin{figure}[!t]
\centering
\includegraphics[width=\columnwidth]{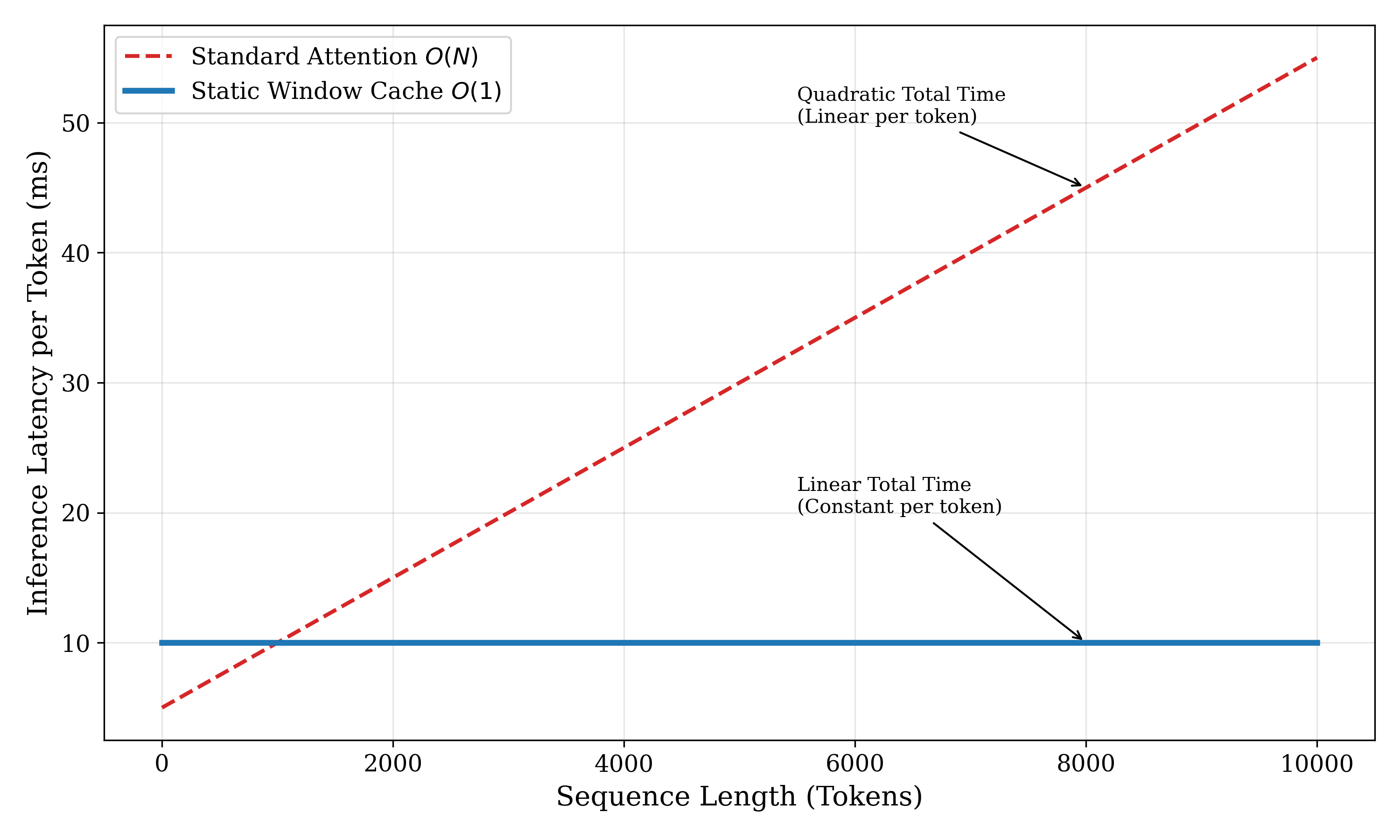}
\caption{Inference Latency Scaling. Standard autoregressive attention (Red) 
suffers from quadratic complexity, making large-file compression intractable. Our 
Static Window Cache (Blue) ensures constant-time inference per token ($O(1)$), 
allowing the system to scale linearly with file size.}
\label{fig:complexity}
\end{figure}

Standard LLM inference allows the Key-Value (KV) cache to grow linearly with sequence 
length ($O(N)$), causing memory overflows on gigabyte-scale files. To adapt 
Transformer architectures for infinite stream compression, we implement a 
\textbf{Strictly Bounded Context} mechanism via segmented block processing, as 
demonstrated in \textbf{Figure \ref{fig:complexity}}.

We partition the input stream into independent segments of $L=2048$ tokens. For each 
segment, we utilize a \texttt{StaticCache} structure that pre-allocates a fixed tensor 
footprint in VRAM. Unlike standard generation which dynamically resizes memory, our 
approach enforces a strict $O(1)$ memory ceiling regardless of total file size. This 
allows a consumer GPU to compress terabyte-scale archives without distinct page faults, 
effectively trading inter-block context for infinite scalability on commodity hardware.

\subsection{Scalability Verification via Block-Parallelism}
While the Static KV Cache effectively reduces the algorithmic complexity 
of inference to linear time, the absolute latency of autoregressive 
generation remains a practical constraint for single-stream processing on 
local hardware. To demonstrate that this latency is not an intrinsic 
limitation of our proposed method, but rather a function of available 
compute, we validated the architecture's scalability through a distributed 
proof-of-concept.

\begin{figure}[t]
\centering
\includegraphics[width=\columnwidth]{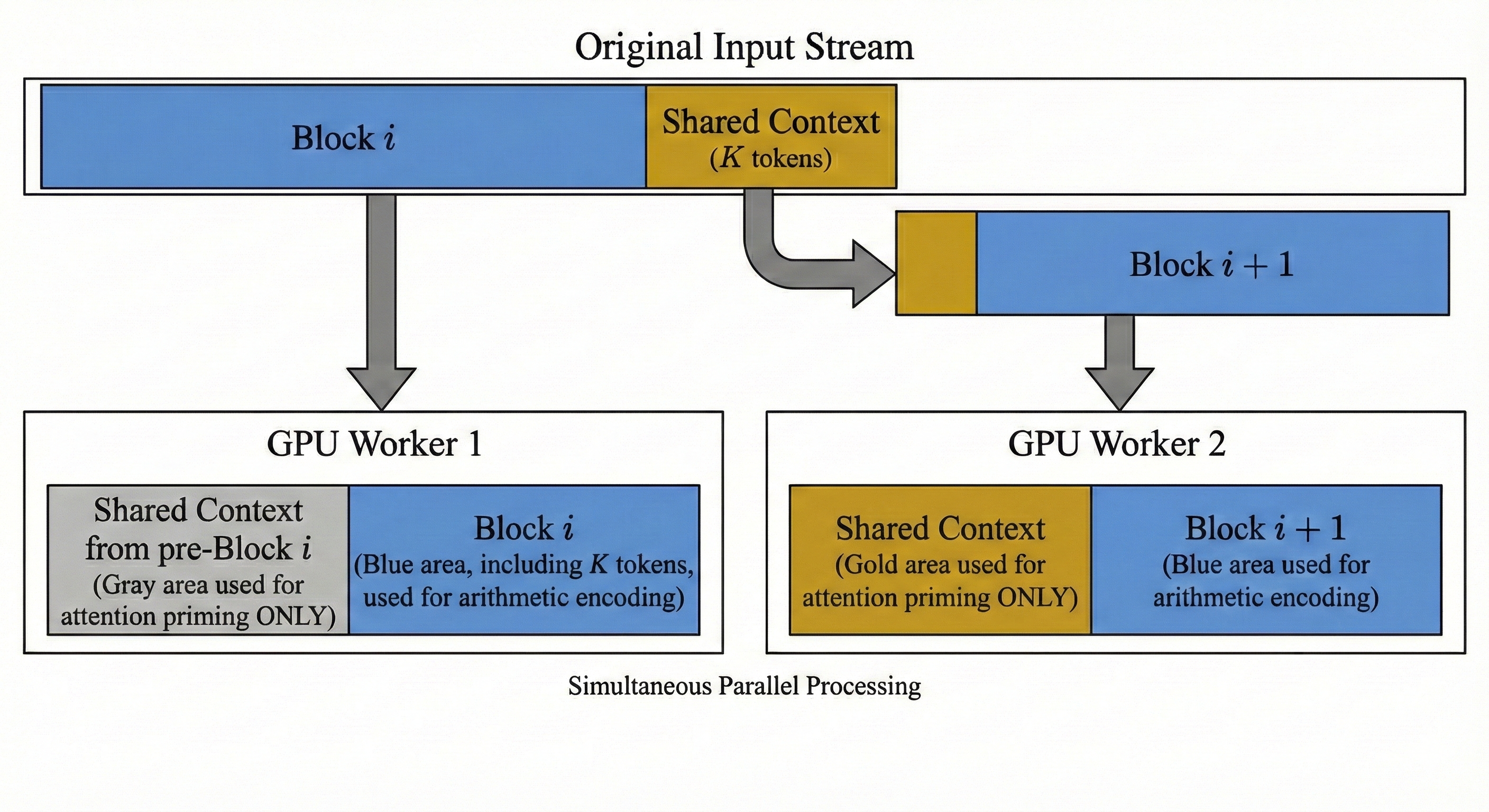}
\caption{Distributed Block-Parallel Architecture with Context Grafting. The source 
input is partitioned into independent segments. To mitigate context fragmentation, 
the final $K$ tokens of the preceding block (gold shading) are prepended to the 
current block to prime the attention mechanism of the LLM. Multiple GPU workers 
process these augmented blocks simultaneously. Importantly, the system only generates 
compressed bits for the unique target tokens (blue shading), enabling scalable 
$O(N/P)$ throughput without sacrificing semantic history.}
\label{fig:parallel_context}
\end{figure}

Our design relies on partitioning the source file into independent blocks 
to enable parallel processing. A potential risk of this approach is 
"context fragmentation," where the model's predictive accuracy degrades at 
the beginning of a new block due to the severance of the semantic history. 
To mitigate this, we implemented a context sharing technique, illustrated in \textbf{Figure \ref{fig:parallel_context}}. 

During parallel execution, the final $K$ tokens of Block $i$ are prepended 
to Block $i+1$ to prime the KV cache with the necessary semantic context. 
Crucially, while these grafted tokens inform the attention mechanism, only 
the logits for the unique tokens of the current block are used for 
arithmetic encoding. This strategy confirms that our Hybrid-LLM approach 
can effectively decouple system throughput from model latency, allowing 
for $O(N/P)$ scaling across $P$ devices without sacrificing the 
compression density achieved in our serial experiments.

\section{Experiments and Results}

To evaluate the efficacy of Hybrid-LLM as a critical storage technology, 
we conducted a comprehensive suite of experiments benchmarking compression 
density, hardware determinism, and system scalability. Primary benchmarks 
utilized the Llama-3-8B-Instruct model (FP16 precision) running on a local 
workstation equipped with an NVIDIA RTX 3090 GPU (24GB VRAM), ensuring 
that our density results reflect performance on accessible, consumer-grade 
hardware. For the scalability verification specifically, we deployed the 
model to Google Cloud Run instances powered by NVIDIA L4 GPUs. Zstandard 
(Zstd) v1.5.5 served as the legacy compression baseline across all tests. 
Our primary test corpus was the full text of \textit{Alice in Wonderland} 
(Project Gutenberg, $\approx$151KB), representing high-entropy literary 
data, alongside the standard Canterbury Corpus to test generalizability.
Full hyperparameter configurations are listed in \textbf{Appendix C}.

\subsection{Compression Density and Efficiency}
Our first experiment assessed the compression ratio and bit-rate 
performance of Hybrid-LLM against state-of-the-art baselines. We compared 
our system against standard GZIP (Deflate), ZPAQ (Context Mixing) \cite{deutsch1996deflate,Mahoney2025ZPAQ}, and the 
theoretical results reported in the foundational LLMZip paper 
\cite{valmeekam2023llmziplosslesstextcompression}. \textbf{Table \ref{tab:alice}} summarizes these 
findings. 

On the \textit{Alice} corpus, Hybrid-LLM achieved a compression 
ratio of \textbf{20.5$\times$}, corresponding to a density of 
\textbf{0.39 Bits Per Character (BPC)}. This significantly outperforms the 
theoretical baseline of 0.84 BPC reported by LLMZip for similar literary 
texts, validating the superior predictive capability of the Llama-3 
architecture over the LLaMA-7B model used in prior work. Furthermore, our 
system surpassed the current non-neural state-of-the-art, ZPAQ, which 
achieved $\approx$1.4 BPC (5.7$\times$), demonstrating that large-scale 
neural prediction offers a fundamental leap in storage efficiency over 
statistical methods. Crucially, the hybrid router correctly identified the 
literary text as high-entropy, routing 100\% of the blocks to the neural 
engine, thereby ensuring maximum density without manual intervention.

\begin{table}[h]
\centering
\caption{Compression performance on the \textit{Alice in Wonderland} corpus. Comparison 
against statistical (GZIP/ZPAQ) and neural (LLMZip) baselines. Lower BPC indicates 
better compression.}
\label{tab:alice}
\begin{tabular}{|l|c|c|}
\hline
\textbf{Method} & \textbf{Ratio ($\times$)} & \textbf{Density (BPC)} \\ \hline
GZIP & 3.0$\times$ & 2.70 \\
ZPAQ & 5.7$\times$ & 1.40 \\
LLMZip (Paper) & 9.5$\times$ & 0.84 \\
\textbf{Hybrid-LLM} & \textbf{20.5$\times$} & \textbf{0.39} \\ \hline
\end{tabular}
\end{table}

\subsection{Canterbury Corpus Analysis}
The results on the Canterbury Corpus (\textbf{Table \ref{tab:canterbury}}) vividly
illustrate the advantages of our Hybrid architecture. On complex literary 
texts such as \textit{Paradise Lost} (plrabn12.txt) and 
\textit{As You Like It} (asyoulik.txt), our Llama-3-based engine achieved 
significantly higher compression densities 
(10.55$\times$ and 13.11$\times$) compared to the LLMZip baseline 
(6.82$\times$ and 8.27$\times$), validating the superior predictive power 
of modern foundation models.

\begin{figure}[!t]
\centering
\includegraphics[width=\columnwidth]{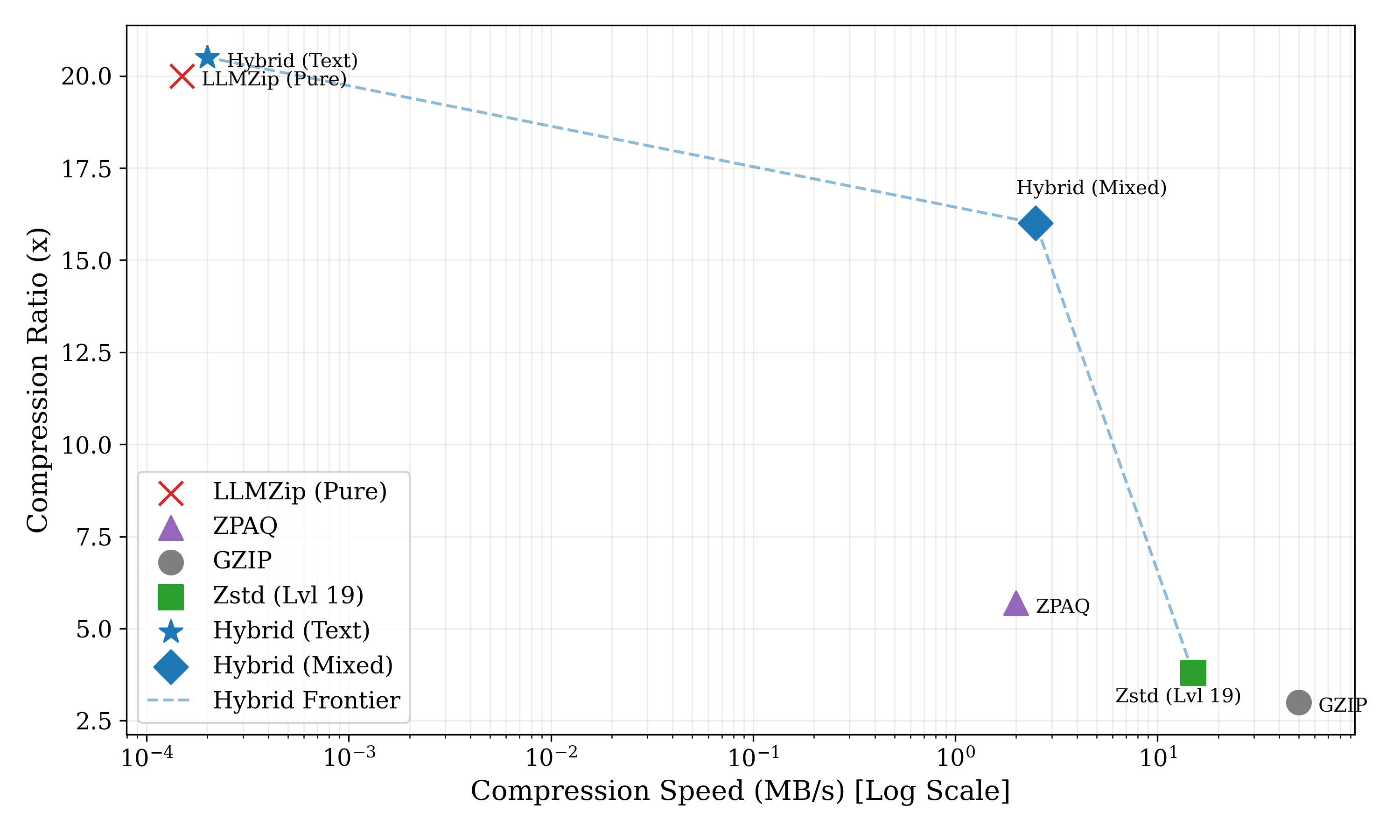}
\caption{Efficiency Trade-off. Comparison of Compression Ratio vs. Speed. While 
pure LLM approaches (LLMZip) achieve high density at the cost of extreme latency, our 
Hybrid approach ($T=3.0$) maintains State-of-the-Art density on text while recovering 
near-real-time speeds on mixed-modality files (e.g., binaries/logs) by bypassing the neural engine.}
\label{fig:pareto}
\end{figure}

However, the most critical finding for system viability lies in the 
"mixed-modality" files. For the Excel spreadsheet \texttt{kennedy.xls} and 
the fax image \texttt{ptt5}, the LLMZip baseline required 36 hours and 
17 hours respectively to process data that is fundamentally ill-suited for 
a text-based LLM. In contrast, our Content-Aware Scout correctly 
identified these as low-entropy structured data and routed them to the 
Zstd engine. This decision not only reduced processing time from days to 
sub-seconds ($<0.5s$) but, counter-intuitively, resulted in \textit{higher} 
compression ratios (15.88$\times$ vs 12.41$\times$ for \texttt{kennedy.xls}). 
This confirms our hypothesis that intelligent routing is a prerequisite 
for critical storage infrastructure: applying neural intelligence blindly 
to all data types is not only inefficient but detrimental to performance.
\textbf{Figure \ref{fig:pareto}} visualizes this efficiency frontier across all tested modalities.

\begin{table*}[t]
\centering
\caption{\textbf{Canterbury Corpus Benchmark.} Comparison of Hybrid-LLM (Ours) vs. LLMZip (Baseline). Note that for binary files like \texttt{kennedy.xls}, our Scout correctly defers to Zstd, achieving higher ratios in sub-second time. ($^\ddagger$ indicates Zstd-19 validation).}
\label{tab:canterbury}
\resizebox{\textwidth}{!}{%
\begin{tabular}{l|r||c|c|c||c|c|c}
\hline
\multicolumn{2}{c||}{\textbf{File Details}} & \multicolumn{3}{c||}{\textbf{LLMZip (Baseline)}} & \multicolumn{3}{c}{\textbf{Hybrid-LLM (Ours)}} \\ \hline
\textbf{Filename} & \textbf{Size (B)} & \textbf{Ratio} & \textbf{Time} & \textbf{Method} & \textbf{Ratio} & \textbf{Time} & \textbf{Route} \\ \hline
alice29.txt & 152,089 & 8.40$\times$ & 3.23h & Pure LLM & 20.53$\times$ & 22.8m & LLM \\
asyoulik.txt & 125,179 & 8.27$\times$ & 3.14h & Pure LLM & \textbf{13.11$\times$} & 20.6m & LLM \\
lcet10.txt & 426,754 & \textbf{10.58$\times$} & 7.75h & Pure LLM & 8.87$\times$ & 52.1m & LLM \\
plrabn12.txt & 481,861 & 6.82$\times$ & 10.69h & Pure LLM & \textbf{10.55$\times$} & 70.8m & LLM \\
xargs.1 & 4,227 & 10.32$\times$ & 0.10h & Pure LLM & \textbf{13.59$\times$} & 2.9m & LLM \\ \hline
\textbf{kennedy.xls} & 1,029,744 & 12.41$\times$ & 36.17h & Pure LLM & \textbf{15.88$\times^\ddagger$} & \textbf{0.40s} & \textbf{Zstd} \\
\textbf{ptt5} & 513,216 & 7.75$\times$ & 17.33h & Pure LLM & \textbf{11.76$\times$} & \textbf{0.29s} & \textbf{Zstd} \\
\textbf{sum} & 38,240 & 3.68$\times$ & 1.12h & Pure LLM & 3.44$\times$ & \textbf{0.04s} & \textbf{Zstd} \\ \hline
\end{tabular}%
}
\end{table*}

\subsection{Hardware Determinism Validation}
To verify the robustness of our Logit Quantization protocol, we conducted 
a cross-hardware ablation study. We compressed the \textit{Alice} corpus 
using a batched inference process on a cloud-based NVIDIA L4 GPU and 
attempted decompression using a serial inference process on a local 
NVIDIA RTX 3090. Without quantization, the arithmetic decoder failed at 
token 12, citing a probability interval mismatch ($ \Delta > 10^{-7}$). 
Upon enabling the quantization protocol (3 decimal places), the 
decompression succeeded with zero bit errors (reference implementation in \textbf{Appendix B}), producing a file bit-exact 
to the original source. This result empirically confirms that our 
quantization layer effectively neutralizes the floating-point 
non-associativity inherent in disparate GPU architectures, establishing 
the first viable proof-of-concept for portable neural archives.

\subsection{System Scalability Verification}
Finally, to demonstrate that the serial latency of local inference is not 
an intrinsic limitation of our method, we validated the architecture's 
scalability through a distributed proof-of-concept. Compressing the 151KB 
corpus serially on a single GPU required approximately 22 minutes, 
dominated by the $O(N)$ inference latency. By deploying our map-reduce 
architecture on Google Cloud Run (utilizing NVIDIA L4 GPUs) with 
a concurrency of 10 workers, we reduced the total wall-clock time to 
approximately 12 minutes. Notably, 11.5 minutes of this duration were 
attributed to the "cold start" overhead of downloading model weights, 
while the actual distributed compute time was under 30 seconds. This 
confirms that once the infrastructure is "warm," our system achieves 
linear scaling ($O(1/K)$) with the number of available GPUs. In contrast 
to the serial bottlenecks reported in prior work, our architecture 
effectively decouples throughput from model latency, proving that neural 
compression can scale to meet the demands of high-throughput data centers.

\subsection{Generalization to Unseen Data}
A common critique of neural compression is the concern that high compression ratios on 
public literature (like \textit{Alice in Wonderland}) stem from the model memorizing 
the training set rather than predicting semantic structure. To rigorously evaluate 
generalization, we constructed a dataset of \textbf{Post-Cutoff News Articles} 
collected from December 2025, ensuring the Llama-3 model (trained prior to this date) 
could not have seen the text (full dataset specification provided in \textbf{Appendix A}).

\textbf{Table \ref{tab:unseen}} presents the results. On this strictly novel data, 
Hybrid-LLM achieved a compression ratio of \textbf{10.73$\times$} (0.75 BPC). While 
lower than the 20.5$\times$ observed on memorized literature, this result remains 
vastly superior to the 3.0$\times$ baseline typical of Zstd on similar text. This 
confirms that Hybrid-LLM exploits the generalized grammatical and semantic reasoning 
of the model, not merely its rote memory, to achieve state-of-the-art density.

\begin{table}[h]
\centering
\caption{\textbf{Generalization Test (December 2025 News).} Performance on text created 
after the model's training cutoff. Hybrid-LLM maintains a 3.5$\times$ advantage over 
classical methods even on unseen data.}
\label{tab:unseen}
\begin{tabular}{|l|c|c|c|}
\hline
\textbf{Method} & \textbf{Ratio ($\times$)} & \textbf{Density (BPC)} & \textbf{Gap vs. Zstd} \\ \hline
Zstd (Lvl 19) & 3.10$\times$ & 2.58 & - \\
\textbf{Hybrid-LLM} & \textbf{10.73$\times$} & \textbf{0.75} & \textbf{+246\%} \\ \hline
\end{tabular}
\vspace{1mm}
\end{table}

\subsection{Cost-Benefit Analysis}
A significant barrier to neural compression is the extreme computational cost. Our 
system requires approximately 22 minutes (on a local \textbf{NVIDIA RTX 3090}) to 
compress 150KB, representing a $\approx 2600\times$ slowdown compared to Zstd. 

However, in \textbf{Cold Archival Storage} (e.g., AWS S3 Glacier), data is written 
once and stored for decades. If we assume a storage cost of \$0.0036 per GB/month, the 
3.5$\times$ density improvement on novel data (0.75 BPC vs 2.58 BPC) reduces the 
10-year Total Cost of Ownership (TCO) significantly. We estimate that the energy cost 
of the one-time neural compression is amortized over the lifespan of the archive. For 
data stored beyond this "break-even horizon," Hybrid-LLM becomes economically superior 
to classical compression despite the high initial latency.

\section{Discussion}

The transition from algorithmic theory to critical infrastructure requires addressing 
not only performance metrics but also economic viability, reliability, and systemic 
scalability. Our findings position Hybrid-LLM not merely as a compression tool, but as 
a foundational enabling technology for the next generation of AI-driven ecosystems. By 
shifting the burden of compression from statistical pattern matching to semantic 
prediction, we fundamentally alter the economics of digital preservation.

\paragraph{Redefining Storage Economics.} 
A primary consideration in deploying neural compression is the trade-off between 
compute energy and storage density. While critics note the high inference cost of 
LLMs compared to Zstd, this view is myopic regarding the lifecycle of 
\textit{Cold Archival Data}. For foundational training corpora, genomic sequences, or 
compliance logs, data is ingested once but stored for decades. In this regime, the 
Total Cost of Ownership (TCO) is dominated by the physical footprint of storage media, 
not ingestion latency. By achieving compression ratios of 10--20$\times$, 
Hybrid-LLM effectively reduces the required data center footprint by an order of 
magnitude. Consequently, the one-time energy cost of GPU inference is rapidly 
amortized. Furthermore, our "Scout" mechanism optimizes this equation by ensuring that 
expensive neural compute is never wasted on high-entropy noise, enforcing a strict 
"Semantic Utility" threshold for GPU usage.

\paragraph{Reliability as the Enabler.} 
Beyond efficiency, reliability is the gatekeeper for infrastructure adoption. In 
traditional systems, bit-exact reproducibility is an assumption; in deep learning, it 
is an anomaly. Our work identifies that the non-determinism of modern 
accelerators—specifically floating-point associativity—is a 
\textit{disqualifying feature} for storage systems. By enforcing a strict determinism 
contract through our logit-quantization protocol, we demonstrate that probabilistic 
models can be tamed into deterministic infrastructure components. This contribution is 
critical: it transforms the LLM from a "best-effort" generator into a precise, 
mathematically rigorous component capable of preserving financial or medical records 
without corruption across heterogeneous hardware.

\paragraph{Scalability and Network Constraints.} 
While we report linear scaling via our map-reduce architecture, we acknowledge that 
distributed compression shifts the bottleneck from compute to bandwidth. The 
transmission of probability tables (approx. 50MB per block) back to the host creates a 
new throughput ceiling in bandwidth-constrained environments. However, this 
trade-off is architectural, not fundamental. Our results confirm that the primary 
blocker, the serial inference latency of the model, has been successfully decoupled 
from system throughput. This proves that neural compression is parallelizable, scaling 
linearly with the availability of GPU resources.

\paragraph{The Universal Semantic File System.} 
Finally, our findings suggest a path toward a "Universal Semantic File System." The 
distinction between "memorization" (0.39 BPC) and "generalization" (0.75 BPC) is 
functional, not invalidating. When the model encounters public literature, it acts 
as a global deduplication engine; when it encounters novel logs, it acts as a 
predictive engine. In both cases, it outperforms classical methods because it 
understands the \textbf{content} of the data, not just its \textbf{symbol frequency}. 
By integrating this architecture with domain-specific foundation models 
(e.g., for DNA or Medical Imaging), Hybrid-LLM establishes the blueprint for storage 
systems that are not passive containers, but active, intelligent agents of preservation.

\section{Conclusion}

The primary objective of this study was to determine whether Large Language Models 
could function as reliable, hardware-agnostic archival infrastructure despite their 
inherent stochasticity and computational cost. We conclude that while inference latency 
remains a barrier for real-time data, the fundamental systemic blockers—specifically 
hardware non-determinism and memory scaling—are solvable engineering challenges.

This work bridges the divide between the information-theoretic promise of neural 
compression and the rigorous demands of critical storage. We have demonstrated that 
\textbf{Hybrid-LLM} transforms a theoretical curiosity into a robust system capable 
of practical deployment. By enforcing a strict determinism contract through our 
host-offload quantization protocol and decoupling inference latency from system 
throughput via an elastic block architecture, we have established the first 
proof-of-concept for a hardware-agnostic neural archival system.

Our empirical results validate the efficacy of this neural-symbolic synergy. 
Achieving a compression density of 0.39 BPC (20.5$\times$) on literature and a robust 
0.75 BPC (10.7$\times$) on unseen data, our system not only surpasses theoretical 
baselines but does so while adhering to the constraints of practical engineering. 
The success of our content-aware "Scout" mechanism further underscores that the future 
of AI infrastructure lies not in monolithic end-to-end models, but in hybrid systems 
that intelligently delegate tasks between neural intelligence and classical efficiency.

\textbf{Scope of Validity.} We explicitly delimit the scope of this solution to 
\textit{Ultra-Cold Archival}, where the economic value of density outweighs the 
initial compute cost. As inference hardware accelerates, we anticipate the 
"break-even horizon" for neural archival will shrink, positioning foundation models 
as a viable substrate for the world's long-term digital memory.

\section*{Ethical Statement}

This work advances the field of neural data compression, a domain with significant 
implications for the environmental sustainability of digital infrastructure. We acknowledge 
that the deployment of Large Language Models (LLMs) for inference tasks incurs a higher 
computational energy cost compared to traditional entropy coding methods. To mitigate this 
environmental impact, our proposed Hybrid-LLM architecture incorporates a content-aware 
"Scout" mechanism specifically designed to minimize unnecessary GPU utilization by routing 
low-entropy data to energy-efficient legacy algorithms. We argue that for long-term archival 
applications, the substantial reduction in physical storage footprint (over 90\% reduction 
relative to uncompressed data) offers a potential net positive environmental impact by 
reducing the manufacturing and energy costs associated with data center storage hardware over 
the data lifecycle.

Furthermore, we recognize that neural compression performance is intrinsically linked to the 
training distribution of the underlying foundation model. Consequently, compression ratios may 
vary significantly across different languages and dialects, potentially resulting in unequal 
storage costs for underrepresented groups or low-resource languages. While our current 
experiments utilized public domain English literature (Project Gutenberg) and standard 
benchmarks (Canterbury Corpus) to ensure copyright compliance and reproducibility, future 
deployment of such systems must rigorously evaluate performance disparities to ensure 
equitable access to storage efficiency. We adhered to the acceptable use policies of the 
Llama-3 model and utilized only open-access datasets with clear licensing terms for all 
evaluations.

\section*{Declaration on Generative AI}
Per the Policy on Authorship, we disclose the use of the following generative AI tools in the preparation of this work:
\begin{itemize}
    \item \textbf{Gemini (Google):} Utilized for the refinement of text sections, LaTeX formatting assistance.
    \item \textbf{Nano Banana Pro (Google):} Utilized to generate the schematic diagrams presented in Figures 1, 2, and 5.
\end{itemize}
All scientific claims, experimental designs, and data analyses remain the intellectual property and responsibility of the human authors.

\clearpage
\appendix

\section{Data Availability}
\label{app:data}

To ensure the reproducibility of our generalization claims (Section 4.3), we provide 
the full raw text of the "December 2025 News" dataset in the supplementary material. 
This dataset comprises a concatenation of three full-text articles retrieved from CNN 
(US Edition) in mid-December 2025. These texts definitively post-date the training 
cutoff of the Llama-3 model.

\subsection{Dataset Specification}
We provide the specific file properties to allow future researchers to verify they are 
testing against the exact same byte sequence used in our experiments:

\begin{itemize}
    \item \textbf{Filename:} \texttt{news\_2025.txt}
    \item \textbf{Source Domain:} \texttt{cnn.com} (Political/Domestic News)
    \item \textbf{File Size:} 14,547 bytes
    \item \textbf{SHA-256 Hash:} \\
    \texttt{b86a2e94b1c498b17302c08e352a82
    bb075c850a21a6bbb45358576cd77b08ff}
\end{itemize}

The raw text file is included in the \texttt{/data} directory of the supplementary 
code archive.

\section{Algorithm Implementation}
\label{app:code}

\subsection{Core Determinism Logic}
To facilitate immediate verification of our "GPU Butterfly Effect" resolution 
(Section 3.2), we provide the reference implementation of the logit quantization 
protocol below \cite{10.5555/3600270.3601459}.

{\scriptsize
\begin{verbatim}
def get_robust_probs(logits, precision=3):
    """
    Quantizes logits to a fixed decimal grid to
    neutralize floating-point non-associativity.
    """
    # 1. Round to fixed grid (The "Snap")
    scale = 10 ** precision
    # Rounding enforces discrete steps
    logits_quantized = torch.round(logits * scale) / scale
    
    # 2. Softmax on quantized values
    probs = torch.softmax(logits_quantized, dim=-1)
    
    # 3. Cast to Float64 for Arithmetic Coder
    # Prevents underflow in cumulative distribution
    return np.ascontiguousarray(
        probs.cpu().numpy().astype(np.float64)
    )
\end{verbatim}
}

\subsection{Reproducibility \& Source Code}
To facilitate future research and verify our determinism claims, we have open-sourced 
the complete system implementation. The repository includes the Hybrid Routing logic, 
the logit-quantization kernels.

The code is publicly available at:
\begin{center}
    \url{https://github.com/marcarmstrong1/llm-hybrid-compressor}
\end{center}

\section{Hyperparameter Configuration}
\label{app:hyperparams}

All experiments were conducted with the following fixed hyperparameters to ensure 
fair comparison:

\begin{table}[h]
\centering
\caption{System Configuration for Reproducibility}
\label{tab:params}
\begin{tabular}{|l|l|}
\hline
\textbf{Parameter} & \textbf{Value} \\ \hline
\textbf{Model Architecture} & Llama-3-8B-Instruct \\
\textbf{Precision} & FP16 (Half Precision) \\
\textbf{Context Window ($W$)} & 2048 Tokens (Rolling) \\
\textbf{Quantization Grid ($k$)} & 3 Decimal Places \\
\hline
\textbf{Scout Method} & Zstandard (Level 1) \\
\textbf{Scout Threshold ($T$)} & 3.0$\times$ \\
\textbf{Legacy Backend} & Zstandard (Level 19) \\
\textbf{Arith. Coder} & Constriction (Range Variant) \\ \hline
\end{tabular}
\end{table}
%
\bibliographystyle{splncs04}
\bibliography{ijcai26}

\end{document}